# Color Image steganography using Deep convolutional Autoencoders based on ResNet architecture

Seyed Hesam Odin Hashemi, Mohammad-Hassan Majidi[*], Saeed Khorashadizadeh

*Abstract*: In this paper, a deep learning color image steganography scheme combining convolutional autoencoders and ResNet architecture is proposed. Traditional steganography methods suffer from some critical defects such as low capacity, security, and robustness. In recent decades, image hiding and image extraction were realized by autoencoder convolutional neural networks to solve the aforementioned challenges. The contribution of this paper is introducing a new scheme for color image steganography inspired by ResNet architecture. The reverse ResNet architecture is utilized to extract the secret image from the stego image. In the proposed method, all images are passed through the prepossess model which is a convolutional deep neural network with the aim of feature extraction. Then, the operational model generates stego and extracted images. In fact, the operational model is an autoencoder based on ResNet structure that produces an image from feature maps. The advantage of proposed structure is identity of models in embedding and extraction phases. The performance of the proposed method is studied using COCO and CelebA datasets. For quantitative comparisons with previous related works, peak signal-to-noise ratio (PSNR), the structural similarity index (SSIM) and hiding capacity are evaluated. The experimental results verify that the proposed scheme performs better than traditional and pervious deep steganography methods. The PSNR and SSIM are more than 40 dB and 0.98, respectively that implies high imperceptibility of the proposed method. Also, this method can hide a color image of the same size in another color image, which can be inferred that the relative capacity of the proposed method is 8 bits per pixel.

## 1. Introduction

With the advent and spread of the Internet of Things (IoT), we have witnessed a complete revolution in the modes and forms of relations and digital media has become the default mode of communications[1]. Nowadays, paper-based memos have been eliminated and documents of the type of digital images have been replaced increasingly. In this new environment and situation, the necessity of secure communication is inevitable[2],[3].

Embedding a secret message within another message such as image, audio, video, or text file is of great importance in secure communication[4]. Steganography studies algorithms in this field to make the message signals undistinguishable while sending to receivers or users through a communication channel[5],[6]. Both steganography and watermarking are types of data hiding. Although watermarking is very similar and close to steganography, there are some considerable differences in their applications. For instance, in watermarking, it is optional to make the message visible or invisible, while in steganography, the message signal must be hidden. Also, in watermarking, the messages hidden in cover are not secret, but in steganography, it is necessary to ensure security firstly[7]. Generally, steganography approaches can be classified into two main categories: transform and spatial domains. In the spatial domain, the cover image data/pixels are directly utilized to hide the secret information. For example, secret bits may be substituted inside cover pixel values. In the transform domain, before applying the embedding process, a conversion of the data of the cover image into other signals/forms is performed at the first stage. Discrete cosine transformation (DCT) is one of these conversions that is applied to cover pixels and then the secret message will be embedded into the different coefficients of DCT blocks [8]. In order to compare the performance of different steganography techniques, some criteria such as payload capacity, stego-image quality, embedding efficiency, and robustness against attacks.

---

[*] **corresponding author:** m.majidi@birjand.ac.ir

In the traditional steganography approaches such as Discrete Wavelet Transformation (DWT), Pixel Value Differencing (PVD), and Least Significant Bits (LSB) substitution, the embedding capacity is limited because of the trade-off between capacity and imperceptibility[9]. Moreover, classical steganography schemes are faced with more extra challenges originating from steganalysis techniques whose main objective is to recover secret information illegally. In order to solve these challenges, deep steganography methods based on convolutional neural networks have been proposed [10].

This paper presents a new architecture for deep steganography that is the main contribution of this paper. At the first step, all images are passed through the prepossess model which is a convolutional deep neural network. This model extracts the features that are used in the embedding and extraction phases. The second step is the operational model in which the extracted features from preprocess model are utilized in order to generate stego and extracted images in the embedding and extraction phases, respectively. In fact, the operational model is an autoencoder based on ResNet structure that produces an image from feature maps. It is worthy to mention that most of existing schemes merge the secret and cover image before their embedding phase that results in an input tensor with 6 channels, while the input tensor of their extraction phase is an image with 3 channels. As a result, the models needed in their embedding and extraction phases are quite different. However, the principal advantage of this scheme that makes it different from the most previous related works is the identity of the models used in embedding and extraction phases. To be more precise, the detail properties of both embedding and extraction phases i.e. The number of layers, types of activation functions and other structural properties are completely the same. As a result, the designer is involved in a simpler trial and error process. This simplicity is due to the preprocess model that merges the feature maps of the secret and cover images. Consequently, the input tensors of the operational model in embedding and extraction phases are the of the same sizes.

The rest of this paper is organized as follows. In Section 2, the related works on steganography-based deep neural networks are introduced. In Section 3, the proposed embedding and extracting phases are described. Section 4 presents the experimental results and compares them with the results of other steganographic methods. Finally, conclusions are presented in Section 5.

## 2. Related works

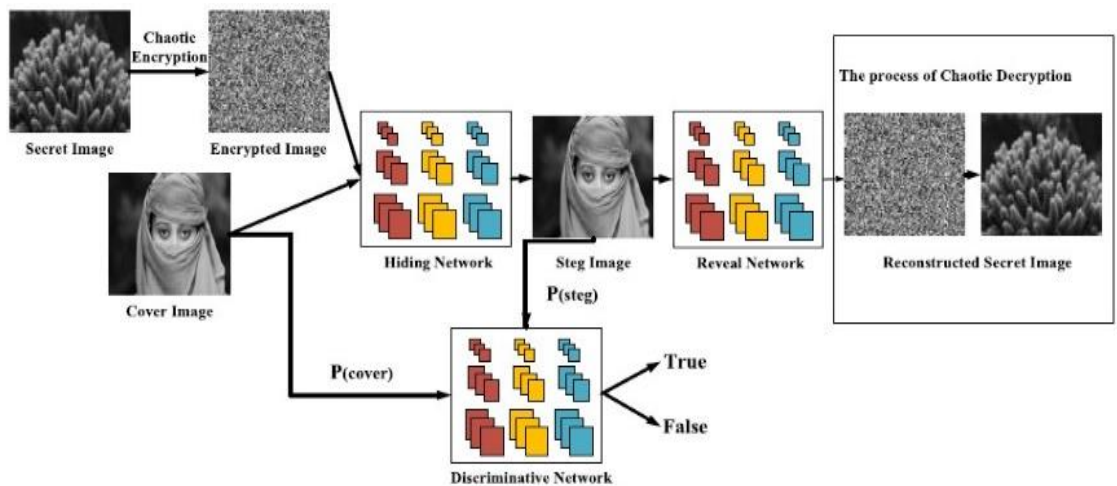

*Fig1: The architecture proposed in [5].*

Recent studies in this field show that there are three different objectives for using deep learning in steganography. The first objective of using deep learning tools is improving the steganography characteristics. In other words, deep learning is used to find the conditions in which the system has a better performance[11], [12]. By way of illustration, a deep network can be used to find the edge or high-frequency parts of images, then the secret is hidden in these parts. As a result, this technique increases the impressibility of traditional methods. As another example, a stereo image reversible data hiding method based on the convolutional neural networks (CNN) has been presented in [13] where a CNN-based predictor is introduced to improve the pixel recovering. In the first stage, pixels are classified into low-resolution (LR) and high-resolution (HR) clusters. The difference view between LR and HR is calculated based on depth and texture information. In the second stage, a CNN is trained to predict the high-precision difference view from the LR difference view, and the high-resolution image is computed to predict pixels accurately. Earlier generative adversarial networks (GAN) are used for information hiding methods. These methods mainly used neural networks as encoder or decoder to transmit the message. However, new methods use GAN with aim of distortion minimization [14]. A steganography scheme based on GAN is proposed in [15]. but his method utilizes the adversarial training of discriminator to improve the quality of stego images for enhancing the security of the system. The higher the visual quality of the stego-image, the less suspicious it becomes, which can increase security. The critic network is designed for the discriminator to evaluate the performance of embedding phase and enhance adversarial training. Also, this model uses two convolutional autoencoders for embedding and recovering images, respectively. In [4], a novel grayscale image steganography scheme based on GAN and chaos encryption has been presented. At the first, the secret image is encrypted by the Chen chaotic system and then, the encrypted secret and cover images are passed through the encoder network which is a convolution autoencoder, and the stego image is generated. Besides, in order to increase the stego image quality further, a GAN is adopted to train the discriminative network using distributions of stego and real images. In the detector, decryption is carried on using the Chen chaotic system and decoder networks implemented by autoencoders are used to reconstruct the secret image from the stego image. Fig 1 shows the architecture of this method.

The second objective of using deep learning in steganography is steganalysis that means revealing the presence of secret messages embedded in digital media illegally. Most methods formulate image steganalysis as a binary classification problem called universal steganalysis. Universal methods are initiated by feature extraction from input images. Then, a binary classifier such as support vector machine (SVM) or ensemble classifier is used to divide images to cover and stego images [16]. In [17], a multi-scale deep neural network model based on the deep residual network for steganalysis is proposed. Also, a wider range of image region correlation extraction is introduced. in order to generate the characteristic channels, some local receptive fields in layers have been used. Various steganographic features were obtained from the scale channels using channel recognition. In comparison with the networks of the single-scale channel, the accuracy of steganalysis is improved using this scheme. In [18], a steganalysis based on the FractalNet model is proposed. FractalNet is a type of convolutional neural network that eschews residual connections in favour of a "fractal" design. In this method, it has been observed that steganalysis for test images is improving if the width of the network can be increased with a certain proportion to the depth. In [19], a multi-class steganalysis technique is proposed. In this method, Residual Network which uses a technique called skip connections was used. Complex statistical features are extracted using a deep residual network. Also, weak stego signal in image content is preserved that is suitable for multi-class steganalysis.

The third objective of using deep learning in steganography is realization of embedding and extraction phases. The approaches presented in this category generally use two autoencoders, one of them is responsible for producing a stego image from cover and secret images, and the other plays the role of recovering the secret image from the stego image. In [20], high-capacity steganography using a deep neural network is introduced. The DCT transform and Elliptic Curve Cryptography (ECC) are employed to transform and encrypt the secret image. In the embedding phase, a collection of hiding and extraction networks in a SegNet deep neural network is used that generates stego images of full-size. In this method, the relative capacity of steganography may reach 1 by effective allocation of each pixel of secret in the cover image. In [21], using deep convolutional autoencoders, an end-to-end image steganography has been proposed. This paper introduces a lightweight yet simple deep convolutional autoencoder architecture. The models include some convolution layers and have simple architectures. Another deep steganography-based deep convolutional network is proposed in [22]. Using the idea of residual blocks in ResNet is the main core of this method. The hiding and extraction procedures are implemented by the encoder and decoder networks in [12]. Both encoder and decoder networks have a similar structure. In a multi-task structure, the aforementioned networks are trained using the same learning rate. However, due to the elimination of the preprocess stage in [12], the required number of epochs for satisfactory training is high. This challenge is solved in the proposed scheme in this paper by applying a suitable preprocess step.

## 3. Proposed method

In this section, the detailed description of each part of the proposed steganography scheme is provided. Fig. 1 illustrates the flowchart of the proposed method in this paper. As shown in this figure, there are two deep neural networks in this scheme. These models are trained simultaneously at each phase. Although they act as a unified network, it is easier to explain their operations and properties separately. The first model, i.e., the preprocess model prepares images to be ready for the second model. The second model that is the operational model, takes the outputs of the preprocess model as inputs to produce an image (stego or extracted image). These two models work in two phases: the embedding and extraction phases. In the embedding phase, secret and cover images are passed through the preprocessing model. Then, they are merged and a feather map with 128 channels is obtained. Next, this feather map is passed through the operational model to generate the stego image. The extraction phase recovers the secret image from the container stego image. During the second phase, stego image is passed through preprocess and operational models with the aim of extracting the secret image that is hidden in the stego image. Let $c, s$ be the cover and secret images, respectively. Producing a stego image $h$ is the main goal of the embedding phase where $h$ is similar to $s$, and, the main aim in the extraction phase is to recover the extracted image $e$ which is similar to $s$. Mathematically, the proposed scheme can be expressed as follows. In embedding phase, $s'$ and $c'$ are feather maps with 64 channels that are generated by the prepresses model ($Prep$).

$$s' = Prep(s), \qquad c' = Prep(c)$$

Then, the operational model ($Om$) produces the stego image $h$.

$$h = Om(s', c')$$

During the extraction phase, $e$ recovers from $h$. In other words,

$$e = Om(Prep(h))$$

More details on each of the models are given in the following subsections.

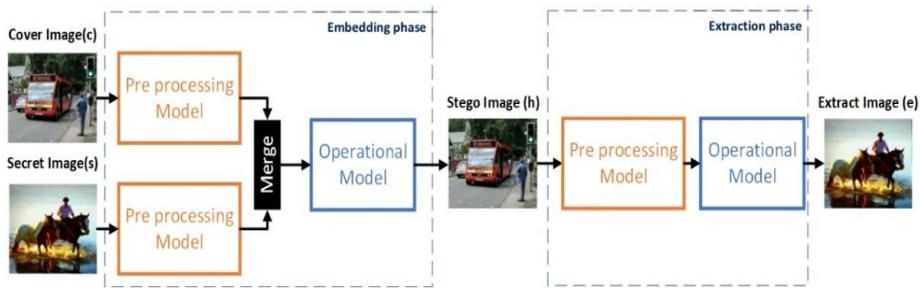

*fig1. The overall block-diagram of the proposed image steganography scheme.*

### A. Prepresses model

Generally, images contain redundant data, and the burden on the embedding and extraction phases will be reduced by extracting the most meaningful features. So, in the embedding and extraction phases, the proposed scheme extracts the primary information from the secret, cover, and stego images as feature maps and prepares them for the operational model. In the both embedding and extraction phases, features produced by preprocessing model are used instead of the raw form of images for steganography. The input size is $M \times N \times O$, which represents the width, height and number of channels of the image. This model contains three convolutional layers with increasing number of filters. The detailed characteristics of the preprocess model are as follows. The kernel is of the size 3×3, the size of stride is 2, and the padding size 1. For initial filters, we should use fewer number of filters to extract lower-level local features, for example the edges of the images. By increasing the number of filters after each layer, the model can find more sophisticated features. There are three convolutional layers in each phase. The number of filters in the first, second and third convolutional layer in the embedding phase is 16, 32, and 64 and in the extraction phase, is 32, 64 and 128, respectively. After each convolution layer, a ReLU and a Batch Normalization (BN) layer is used. The reason of using the ReLU layer is increasing the nonlinear fitting property of the model. As a result, it will learn the details of the secret image, cover, and stego images better. Also, the batch normalization layer is used (except for the first and last convolutional layers) to speed up the network. This model is like an encoder in an autoencoder that produces a feature map from an image.

### B. Operational model

Similar to the preprocess model, the operational model is used in both the embedding and extraction phases. In the embedding phase, the two images passed through the preprocess model are merged and a feature map with 128 channels is generated. Also, the feature map that is produced in the preprocessing model of the extraction phase, has 128 channels. So, the input tensor of the operational model in both the embedding and extraction phases has 128 channels. It is obvious that the output size is always $M \times N \times O$. In the embedding phase, this model generates a stego image ($s$) and during the extraction phase, this model generates the extracted image ($e$). Convolutional neural networks are one of the best-known architectures for autoencoders in encoder and decoder. But when the depth of CNN architecture increases, the problem of vanishing descent arises which affects the accuracy. In this paper, the aforementioned challenge is addressed using the residual block connection technique which is known as ResNet. Utilizing this connection method, the rate of reusing the features is increased. Consequently, the imperceptibility of the stego images is enhanced, and finally, in the extraction phase, the quality of extracted images is improved. In other words, the similarity between the cover and stego images and also between the secret and extracted images will be increased. Fig. 2 presents the idea of feature connection in ResNet. The symbol "⊕" in this figure, means the element-level addition that is known as shortcut. It should be noticed that the two feature layers $F(x)$ and $x$ must have the same size. The idea of cross-layer connection has been applied in Resnet and nowadays, and we have witnessed widespread applications of ResNet as a popular classification model.

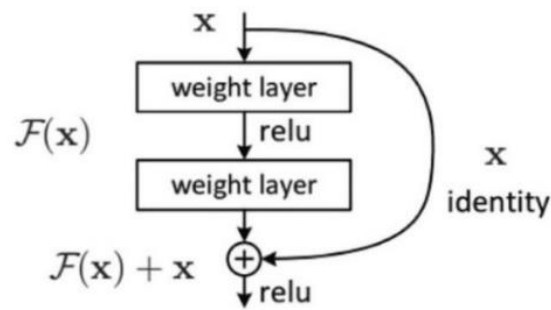

*Fig2. Residual block of ResNet*[22].

The operational model is contained three residual block connections, including 7 convolution transpose operations, Batch normalization (BN) and Leaky-ReLU layers. These layers work like a decoder in an autoencoder and reproduce an image from a feature map. The detailed characteristics of the convolution transpose operations are as follows. The kernel is of the size 3×3 and 4×4, the size of stride is 2 and 1, and the padding size 1.

In this scheme, the input of the extracting phase is the output of the embedding phase, and this connection between the two phases during the training is simplified the generation of the

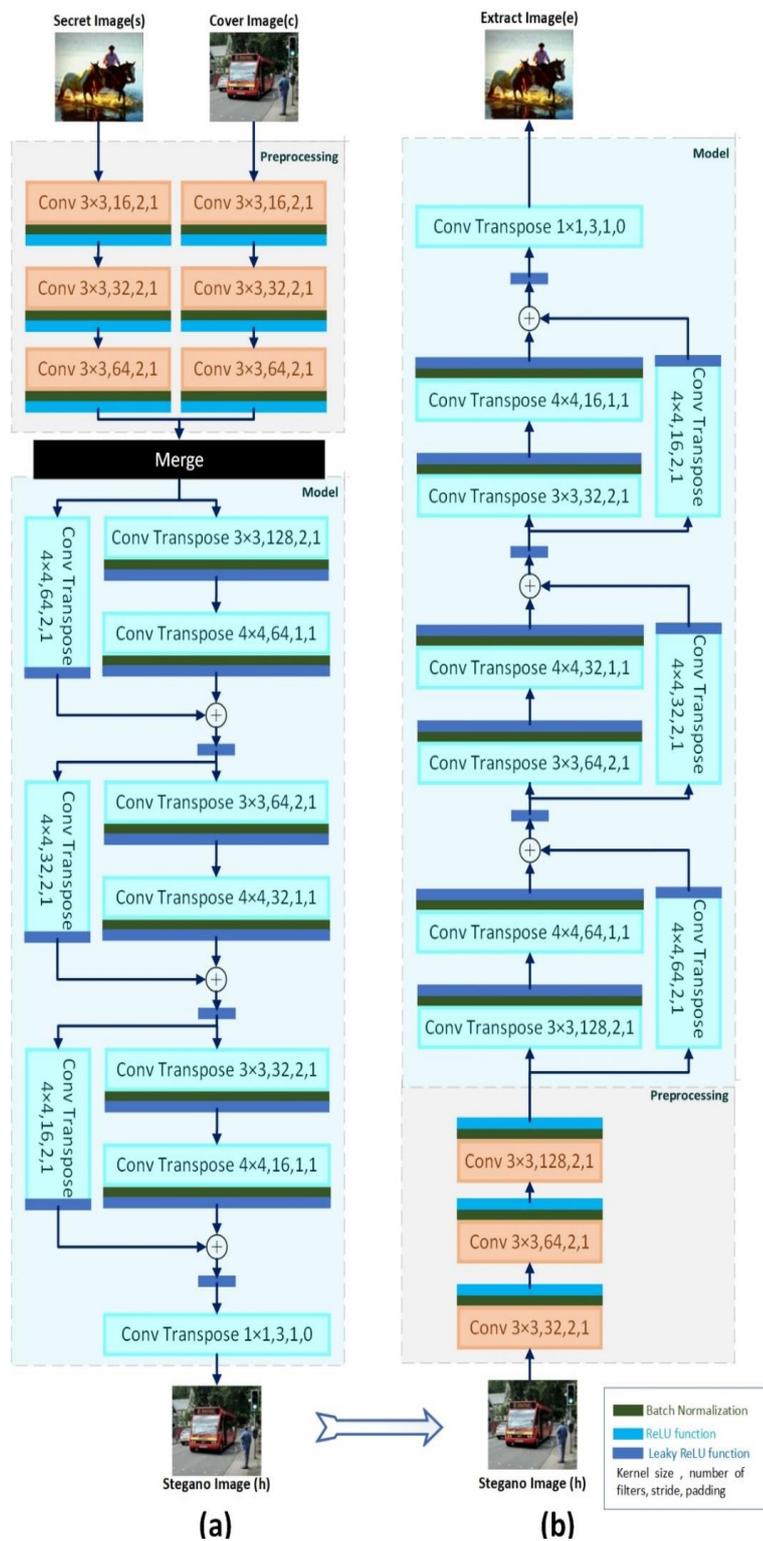

*Fig3. the architecture of the proposed method. a) Embedding phase. b) extraction phase*

extracted images. In the other words, at the first the stego image is produced by the embedding phase, then this image is passed through the extraction phase, and is produced extracted image. So, all networks are learned their parameters at each iteration together. Fig. 3 represents the architecture of the proposed method.

### C. Loss function

The higher the visual quality of the stego-image, the less suspicious it becomes, which can increase security. Also, the extracted image must be similar to the secret image. So, we selected mean square error (MSE) to measure the similarity between the two images. MSE is calculated as:

$$MSE(i, i') = \frac{1}{M \times N \times O} \sum_{c=1}^{O} \sum_{n=1}^{N} \sum_{m=1}^{M} (i_{m,n,c} - i'_{m,n,c})^2$$

where $i_{m,n,c}$ is the pixel $(m, n)$ of image $i$ in channel $c$. Finally, the overall loss is calculated by:

$$Loss = \alpha \times MSE(c, h) + (1 - \alpha) \times MSE(s, e)$$

with the hyper-parameter $\alpha$, the designer can trade off the visual quality of stego image against the extracted images. If the higher α is selected, the higher the visual quality of the stego image is achieved. In contrast, if the lower α is selected, the higher quality of the extracted image is restored. Because of the differentiable functions, the system can be trained end-to-end.

### 4. Experimental results and discussions

In this section, the numerical simulation experiments are presented to verify satisfactory performance of the proposed scheme. The proposed method was implemented in Python3 with

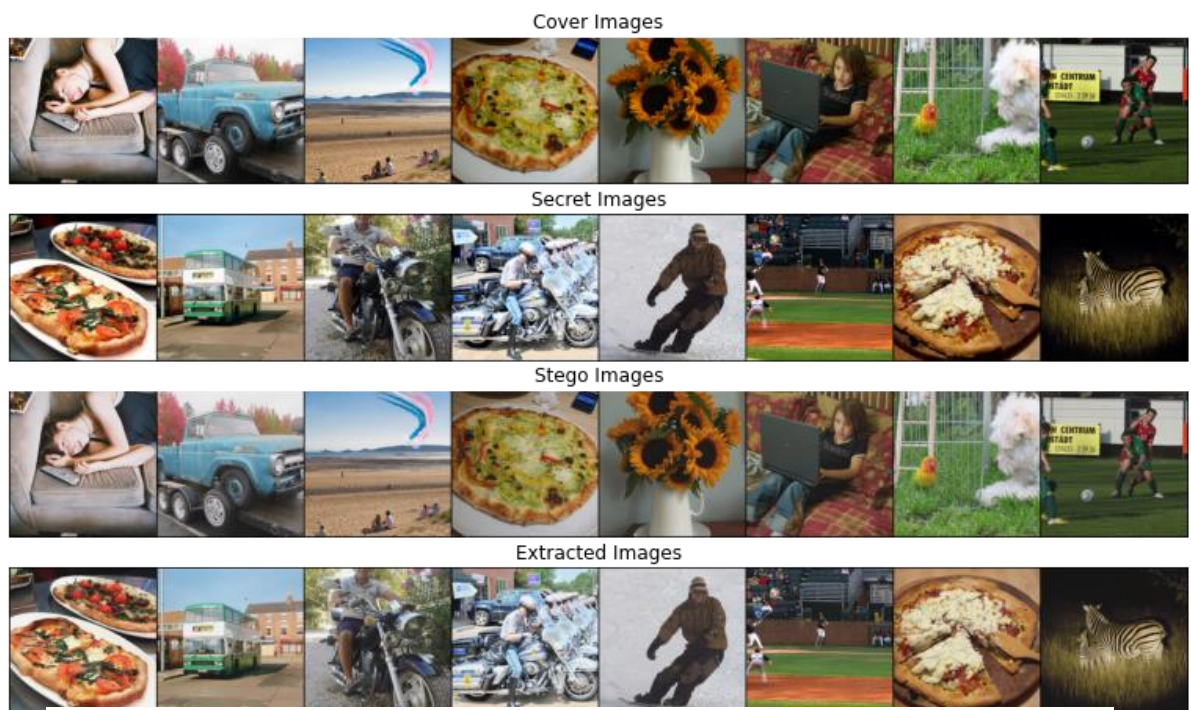

*(a)*

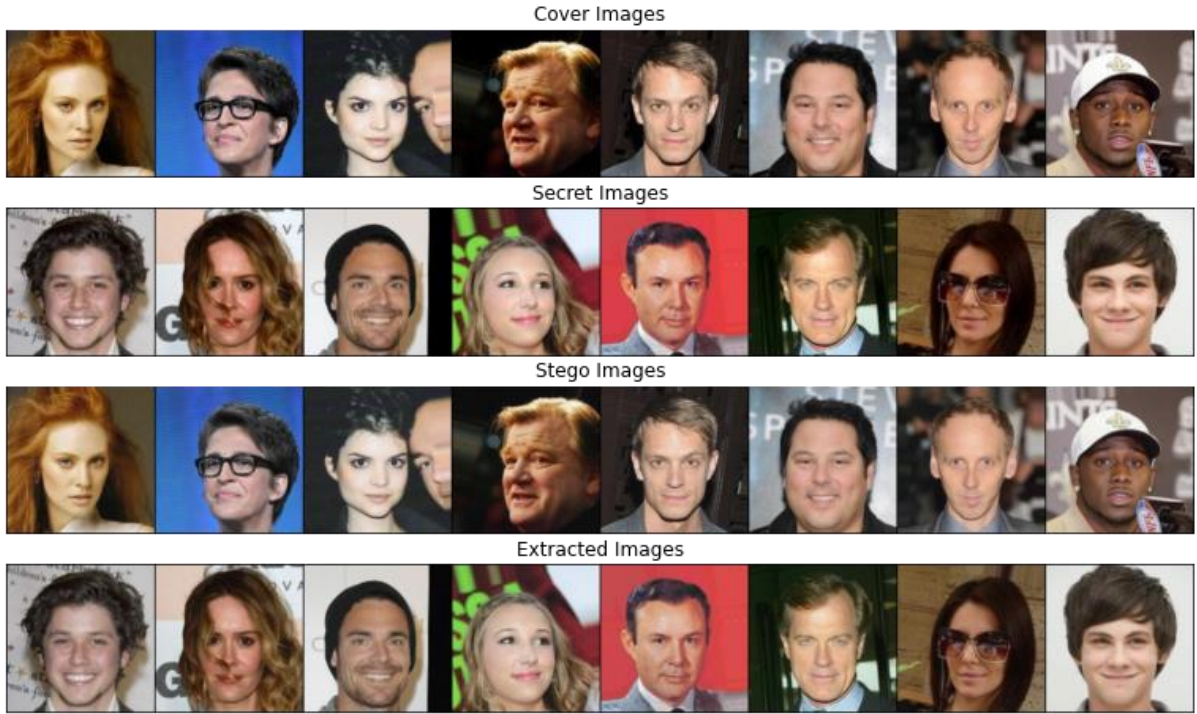

*(b)*

*Fig4. Some examples of the proposed method's results. (a) Dataset COCO and (b) celebA*

PyTorch framework on the NVIDIA GeForce RTX 3090 Graphics Processing Unit (GPU). We trained our models on two datasets: CelebA [23] and COCO [24]. CelebA contains 202K large scale face images. For COCO dataset, we used the test collection of this dataset which has 41K images from general objects. We selected randomly 15K of images from datasets and the learning rate of the training system was 0.001. All the images in the two datasets are resized to 256×256, then images are split in two groups randomly for cover and secret images. The models are trained with Adam optimizer and batch size of 100. This system is trained for 2000 epochs, and each epoch averagely spent 300 seconds for training.

Fig. 4 shows the performance of the proposed method for some images. The cover, secret, stego, and extracted images are shown in rows 1,2,3, and 4, respectively. It can be seen from the images, that the proposed method efficiently can hide the secret image in the cover image and produce a stego image, and in the receiver, the scheme can extract the secret image from

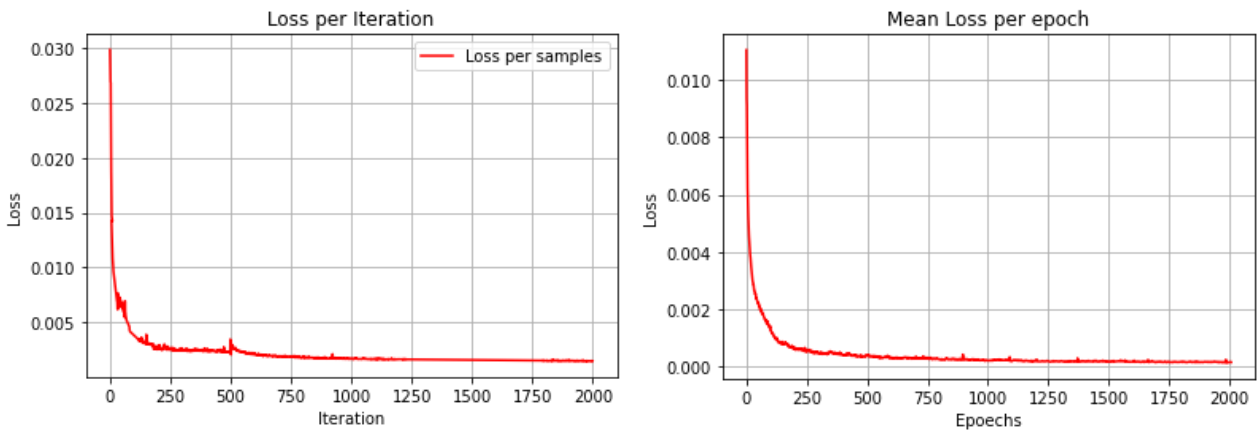

*(a)*          *(b)*

*Fig 5. Mean Loss per epoch. (a) COCO dataset. (b) CelebA dataset*

the stego image with high similarity. Fig 5. Shows the decreasing average loss function for all images per epoch during the training step.

In order to analyze the efficiency of the proposed method, the visual quality of images, payload capacity, hidden capacity and steganalysis are presented in this section.

### *4.1 Imperceptibility*

In general, the stego image must be similar to the cover image and the extracted image must be similar to the secret image. The results of visual effects of the proposed method are measured by the quality of the imperceptibility using PSNR and SSIM. In the case of image steganography, the secret image plays the role of destructive noise, which reduces the quality of the stego image. The steganography algorithm is said to have a good performance, if the PSNR value between cover and stego image are greater than 35 dB and SSIM value must be close to 1. So, higher PSNR and SSIM values between cover and stego image implies smaller distortion after steganography.

PSNR results from the calculation of the logarithm of MSE of two images. The formula for PSNR is as follows:

$$PSNR(i, i') = 10 \times \text{Log}_{10}(\frac{Max^2}{MSE(i, i')})$$

where $Max$ represents the maximum pixel value of the image. Forasmuch as RGB images have 8 bits per pixel, the $Max$ value is 255.

SSIM is another measure of the similarity of two images that uses human perception and structural features of images. This criterion is calculated by three principal coefficients: contrast, luminance, and structure. The formula is as follows:

$$SSIM = l(i, i') \times c(i, i') \times s(i, i')$$

$$l(i, i') = \frac{2\mu_i \mu_{i'} + C_1}{\mu_i^2 + \mu_{i'}^2 + C_1}$$

$$c(i, i') = \frac{2\sigma_i \sigma_{i'} + C_2}{\sigma_i^2 + \sigma_{i'}^2 + C_2}$$

$$s(i, i') = \frac{\sigma_{ii'} + C_3}{\sigma_i \sigma_{i'} + C_3}$$

The first coefficient is $l(i, i')$, a function that compares luminance between two images. The second coefficient compares the contrast of two images, which is notated by $c(i, i')$. The last coefficient is $s(i, i')$ that compares structures of two images. Also, $\mu_i$ and $\sigma_i$ are mean and standard deviation pixels of an image, respectively and $\sigma_{ii'}$ is a covariance between image $i$ and image $i'$. The parameters $C_1, C_2$, and $C_3$ are constant values to avoid the zero denominators. According to the previous studies, the values $C_1 = (0.01 \times 255)^2, C_2 = (0.03 \times 255)^2$, and $C_3 = \frac{C_2}{2}$ are recommended as the default values [25]. The range of SSIM is $(-1,1)$, when two images were exactly the same, SSIM takes the value of 1.

The PSNR and SSIM values of COCO and CelebA datasets were calculated for the proposed method. Tables 1 shows the average PSNR and SSIM values of the proposed scheme for 100 images that are selected randomly from datasets and results of other steganography methods.

*Table 1. PSNR and SSIM values.*

|  | cover and stego images | | secret and extracted images | |
|---|---|---|---|---|
|  | **PSNR value** | **SSIM value** | **PSNR value** | **SSIM value** |
| Traditional Methods | | | | |
| **1-Bit LSB method** | 59.769 | 0.9993 | 29.118 | 0.3277 |
| **2-Bit LSB method** | 52.577 | 0.9963 | 29.566 | 0.6730 |
| **4-Bit LSB method** | 40.498 | 0.9462 | 33.729 | 0.9578 |
| **7-Bit LSB method** | 35.118 | 0.2199 | 55.884 | 0.9994 |
| **Method [26]** | 34.159 | 0.9549 | 39.254 | 0.9615 |
| **Method [27]** | 36.268 | - | 35.981 | - |
| Deep Steganography Methods | | | | |
| **Method [22] ($\alpha = 0.3$)** | 39.561 | 0.9743 | 42.325 | 0.9920 |
| **Method [22] ($\alpha = 0.5$)** | 40.228 | 0.9810 | 40.313 | 0.9899 |
| **Method [22] ($\alpha = 0.7$)** | 41.247 | 0.9839 | 40.062 | 0.9852 |
| **Method [1] COCO Dataset** | 31.960 | - | 27.900 | - |
| **Method [1] CelebA Dataset** | 32.260 | - | 27.920 | - |
| **Method [28]** | 40.57 | 0.9683 | - | - |
| **Method [15]** | 36.41 | 0.9564 | 43.752 | 0.9884 |
| Deep Steganography Proposed Method | | | | |
| **CelebA Dataset ($\alpha = 0.5$)** | 41.0988 | 0.9908 | 40.7949 | 0.9897 |
| **COCO Dataset ($\alpha = 0.5$)** | 39.0617 | 0.9885 | 35.810 | 0.9834 |

The scatter plots presented in Fig.6 show the maximum, minimum and average values of the PSNR or SSIM scheme with $\alpha$ = 0.5, as well as the distribution status of all test data. Using scatter plots to display the test set PSNR and SSIM facilitates analysis and comparison huge data.

These results indicate that the PSNR values are above 40dB and the SSIM value is greater than 0.98. Therefore, the results verify good imperceptibility of the proposed method and secret image cannot be detected by the human visual system.

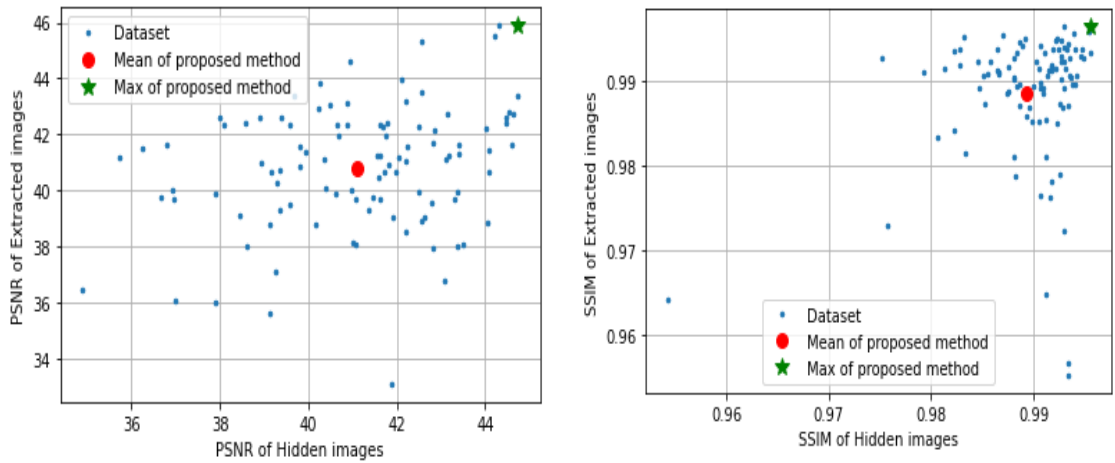

*Fig 6. The scatter plots of PSNR and SSIM.*

## 4.2 Capacity

The amount of secret information that can be hidden without affecting the visual quality of cover image is called "relative capacity". Relative capacity is computed by the hidden information per pixel of the steganography system. Because the cover and secret images in this method are the same size of ($256 \times 256 \times 3$), the capacity of this scheme is 8 bits per pixel. Generally, the deep steganography methods have higher capacity than the traditional arterially designed embedded algorithms. Relative capacity is calculated as follows:

$$Relative\ capacity\ (bpp) = \frac{Absolute\ capacity}{M \times N}$$

We also calculate the payload capacity of information of the image in the proposed method by the following formula.

$$payload\ capacity\ (bpp) = (1 - \frac{\sum_{i=1}^{N}\sum_{j=1}^{M}|S_{i,j} - R_{i,j}|}{N \times M}) \times 8 \times 3$$

The number of bits of information accurately contained in each pixel is referred to as payload capacity. Let $N$ and $M$, the width and height of the image, respectively, and $S$ and $R$ denote the secret image and the extracted image, respectively.

Table 2 compares the payload and relative capacities between the proposed and other methods. According to Table 2, our proposed method has a much higher steganographic capacity than several other methods. This is also a significant advantage of deep convolutional neural network-based image steganography.

*Table 2- Comparison of payload and hiding capasites between the proposed method and other methods.*

|  | Payload Capacity (bpp) | Relative capacity (bpp) |
|---|---|---|
| **1-Bit LSB method** | 21.771 | 1 |
| **2-Bit LSB method** | 22.911 | 2 |
| **4-Bit LSB method** | 23.054 | 4 |
| **7-Bit LSB method** | 23.984 | 7 |
| **Method [20]** | 4.687 | $9.16 \times 10^{-3}$ |
| **Method [29]** | 0.153 | $6.40 \times 10^{-3}$ |
| **Method [30]** | 0.0963 | $4.0 \times 10^{-3}$ |
| **Method [31]** | 22.520 | 0.280 |
| **Proposed method (CelebA)** | 23.608 | 8 |
| **Proposed method (COCO)** | 23.426 | 8 |

## 4.3 Steganalysis

In this section, the security and reliability of the proposed method are analyzed in three ways which are histogram, differential, and StegExpose. To achieve high security, the cover and stego images should not be classified correctly by histograms. The first and second row in Fig. 7 represent histogram analysis for cover and stego images, respectively. As it can be seen in this figure, the difference between cover and stego images is much small, and this can indicate that the proposed method does not contain additional noise very much. Thus, it is difficult to recognize the hidden information after an image has been modified using the proposed method.

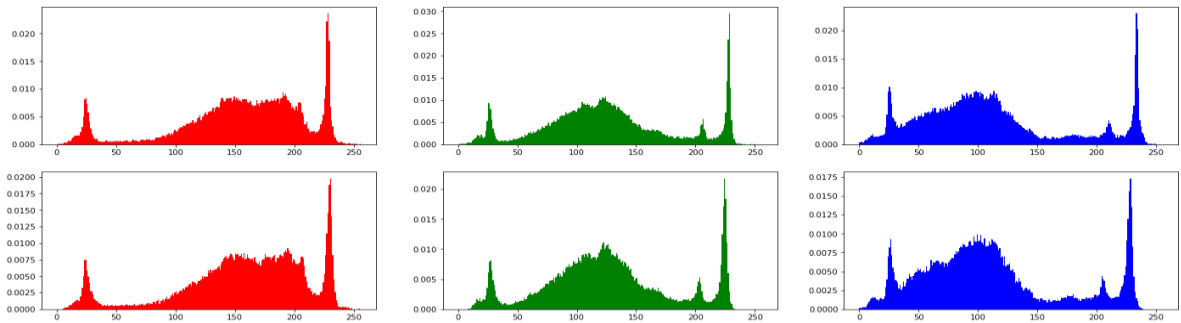

*Fig 7. Histogram analysis: the first row shows the histogram of cover image and the secend row shows the histogram of stego image for 3 channels red, green and blue.*

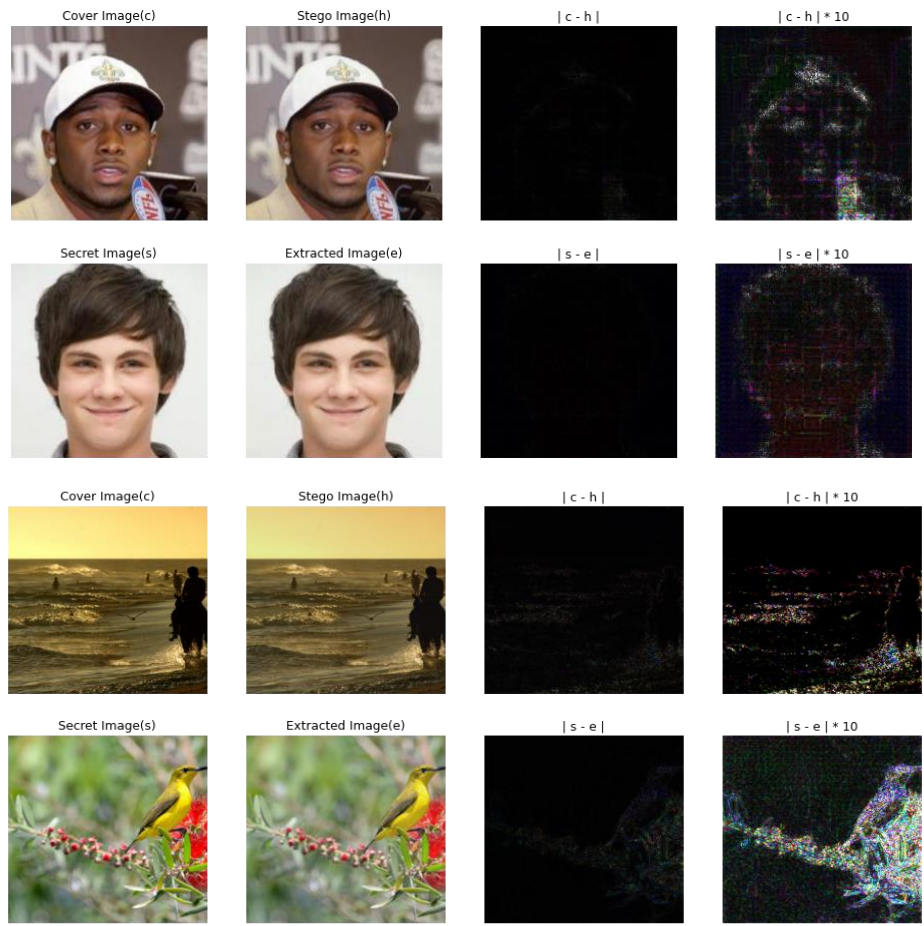

*Fig 8. The difference analysis.*

Fig 8. represents the difference analysis in the proposed method. Column three shows the difference between the cover and stego images, and the secret and stego images, respectively. Also, column four shows the 10 times amplified of the difference. As shown in this figure, an analyzer cannot extract any information from the difference between stego and cover images, or secret and extracted images.

## 5. Conclusion

In this paper, a deep learning-based color image steganography scheme by combining convolutional autoencoders and ResNet architecture has been proposed. Traditional steganography methods suffer from some critical defects such as low capacity, security and robustness. In recent decades, image hiding and image extraction were realized by autoencoder convolutional neural networks to solve the aforementioned challenges. In the proposed method, all images are passed through the prepossess model which is a convolutional deep neural network with the aim of feature extraction. Then, the operational model generates stego and extracted images. In fact, the operational model is an autoencoder based on ResNet structure that produces an image from feature maps. The advantage of proposed structure is identity of models in embedding and extraction phases. The performance of the proposed method was studied using COCO and CelebA datasets. For quantitative comparisons with pervious related works, PSNR, the SSIM and hiding capacity are evaluated. The experimental results verify that the proposed scheme outperforms.